\documentclass[submission,copyright,creativecommons]{eptcs}
\providecommand{\event}{LSFA 2026}

\usepackage{iftex}
\ifPDFTeX
  \usepackage{underscore}
  \usepackage[T1]{fontenc}
  \usepackage[utf8]{inputenc}
\fi
\newcommand{\imply}{\supset}

\usepackage{amsmath,amssymb}
\usepackage{booktabs}
\usepackage{array}
\usepackage{tikz}
\usetikzlibrary{arrows.meta,positioning}

\usepackage{url}

\usepackage[T1]{fontenc}
\usepackage{graphicx}
\usepackage{bussproofs}
\EnableBpAbbreviations%

\newcommand*\Rlbl[1]{{\RL{\footnotesize{#1}}}}
\usepackage{amsmath}
\usepackage{bbold}
\usepackage{amsthm}
\usepackage{adjustbox}
\usepackage{tikz}
\usetikzlibrary{shapes.geometric}
\usetikzlibrary{quotes}
\usepackage{float}
\usepackage{circuitikz}
\usetikzlibrary{automata}
\usetikzlibrary{calc}
\usetikzlibrary{positioning}
\usetikzlibrary{circuits.logic.US,arrows} 
\usepackage[font=small]{caption}
\usetikzlibrary{quantikz2}
\usepackage{comment}
\usepackage{hyperref}
\usepackage{listings}
\usepackage{float}
\usepackage[most]{tcolorbox}
\usetikzlibrary{arrows.meta,decorations.pathmorphing,positioning}
\usetikzlibrary{backgrounds}
\usepackage{changepage}
\usepackage{subcaption}
\usepackage{placeins}

\usepackage{listings}
\usepackage{xcolor}
\usepackage{courier} 
\newtheorem{theorem}{Theorem}
\newtheorem{definition}[theorem]{Definition}

\lstdefinelanguage{Lean}{
  morekeywords={
    def, theorem, lemma, example, structure, inductive, namespace, end, open,
    variable, variables, section, parameter, parameters, axioms, axiom, theorem,
    where, match, with, by, have, show, from, fun, if, then, else, do, let,
    Type, Prop, intro, elim, rewriting, rewrite, simp, exact, refine, apply,
    assume, forall, exists, sorry, repetition
  },
  sensitive=true,
  morecomment=[l]--,
  morecomment=[s]{/-}{-/},
  morestring=[b]"
}

\lstdefinestyle{leanstyle}{
  language=Lean,
  basicstyle=\linespread{0.88}\ttfamily\footnotesize,
  keywordstyle=\bfseries\color{blue!60!black},
  commentstyle=\itshape\color{green!40!black},
  stringstyle=\color{red!60!black},
  columns=fullflexible,
  keepspaces=true,
  showstringspaces=false,
  numbersep=8pt,
  numbers=left,
  numberstyle=\tiny\color{gray},
  frame=single,
  rulecolor=\color{black!15},
  framerule=0.3pt,
  breaklines=true,
  tabsize=2,
  upquote=true,
  extendedchars=false,
  inputencoding=utf8,
}

\newcommand{\gettikzxy}[3]{%
\tikz@scan@one@point\pgfutil@firstofone#1\relax
  \edef#2{\the\pgf@x}%
  \edef#3{\the\pgf@y}%
}
\newcommand*\opi[1]{\mathop{\mathit{#1}}}
\newcommand*{\HC}{\opi{HC}}
\newcommand*\R[1]{\ensuremath{\mathbb{#1}}}

\usetikzlibrary{circuits.logic.US, positioning, calc, arrows}

\title{From Dag-Like Proofs to Boolean Circuits in Lean}

\author{
Lorenzo Saraiva
\institute{Departamento de Inform\'atica\\
Pontif\'icia Universidade Cat\'olica do Rio de Janeiro (PUC-Rio)\\
Rio de Janeiro, Brazil}
\email{lsaraiva@inf.puc-rio.br}
\and
Edward Hermann Haeusler
\institute{Departamento de Inform\'atica\\
Pontif\'icia Universidade Cat\'olica do Rio de Janeiro (PUC-Rio)\\
Rio de Janeiro, Brazil}
\email{hermann@inf.puc-rio.br}
}

\def\titlerunning{From Dag-Like Proofs to Boolean Circuits in Lean}
\def\authorrunning{L. Saraiva \& E.H. Haeusler}

\begin{document}
\maketitle

\begin{abstract}
In this article, we present a method for encoding Dag-Like Derivability Structures (\textbf{DLDS}), obtained via horizontal compression of Natural Deduction proofs in purely implicational minimal logic ($M_{\imply}$), as Boolean circuits. These \textbf{DLDS} compress Natural Deduction tree-like proofs into directed acyclic graphs, preserving logical correctness while reducing redundancy. We formally define the circuit construction process and establish its pointwise correctness, showing that, for any fixed path assignment, the resulting Boolean circuit agrees with the intended dependency-propagation semantics. A Lean formalization establishes machine-checked guarantees for the circuit evaluator and includes a restricted bridge for the uncompressed simple-tree fragment, connecting valid \textbf{DLDS} instances in that fragment to genuine circuit acceptance of their extracted paths under the route and discharge conditions formalized in Lean. This approach opens new perspectives for automated theorem proving and formal certification.
\end{abstract}

\section{Introduction}

Compressed proof representations are essential for efficient proof checking in automated theorem proving and formal verification. The purely implicational fragment of minimal logic ($\mathsf{M}_{\supset}$) provides an ideal setting for this work: its provability problem is PSPACE-complete and polynomially simulates both classical and intuitionistic propositional logic. Results in this system are therefore representative of a broad class of propositional logics from a complexity-theoretic standpoint.

Dag-Like Derivability Structures (\textbf{DLDS})~\cite{haeusler2025horizontal}, obtained via horizontal compression of Natural Deduction (\textbf{ND}) proofs, reduce proof size while preserving logical correctness. The Horizontal Compression (\textbf{HC}) algorithm transforms a tree-like \textbf{ND} proof $\Pi$ into a compressed \textbf{DLDS} $\HC(\Pi)$ by identifying and collapsing identical formulas at the same derivation level. The algorithm runs in polynomial time in the size of the proof, producing a rooted graph where the number of nodes is bounded by $\opi{size}^{2}(\alpha)$ for conclusion formula $\alpha$. The \textbf{DLDS} $\HC(\Pi)$ is valid if and only if $\Pi$ is a closed proof with all assumptions discharged. \textbf{DLDS} validity can be verified in polynomial time via the Flow function, which propagates dependencies through the compressed graph structure. However, this verification is tightly coupled to the specific topology of each \textbf{DLDS}. The Flow-based approach does not expose the verification condition as an explicit, uniform Boolean function that could be evaluated independently of the \textbf{DLDS} structure or reasoned about using alternative verification paradigms.

We reformulate the dependency-propagation check associated with a \textbf{DLDS}
as a uniform Boolean evaluation problem \(f_{\mathcal D}\) over a fixed input
domain, and show that \(f_{\mathcal D}\) is computed by a polynomial-size
Boolean circuit built from the \textbf{DLDS}. The reformulation decouples verification from the underlying graph:
rather than traversing a particular structure, one evaluates a canonical Boolean
function whose inputs encode candidate derivation paths. We are explicit about what
this does \emph{not} claim. In the classical setting it does not improve on Flow:
the global acceptance condition is a conjunction over an exponential space of path
assignments and is not a computable object. The genuine classical content is
\emph{pointwise}: correctness for an arbitrary fixed path assignment, from which
the global characterization follows by universal quantification, as a definitional
move. The value of the reformulation lies elsewhere: a uniform Boolean encoding of
proof validity is amenable to verification paradigms that operate on Boolean
functions rather than on graph structure. In particular, deciding whether the
circuit computes the constant-$1$ function is a global property of a Boolean
function, a setting in which quantum techniques such as amplitude amplification can
potentially be applied. We treat the design of concrete quantum procedures as
future work (Section~\ref{future}); the present article concerns the classical
construction and its mechanized correctness.

Boolean circuits have precise operational semantics and are well suited to mechanization in interactive theorem provers. We use the Lean proof assistant~\cite{Lean} to formalize the translation from \textbf{DLDS} to circuit and to establish its correctness. The development proceeds from the semantics of individual subcircuits to the evaluation of the whole circuit, culminating in the pointwise correctness theorem and the global condition obtained from it. To our knowledge this is the first mechanized treatment of \textbf{DLDS}-to-circuit encoding in any proof assistant.

\textbf{Contributions.} Our main contributions are:
\begin{enumerate}
\item A reformulation of \textbf{DLDS} verification as a uniform Boolean
evaluation problem over path assignments, recasting the checking of candidate
dependency-propagation paths as evaluation over a fixed input domain rather
than as direct graph traversal (Section~\ref{sec:Boolean}).

\item A polynomial-size Boolean circuit construction for this evaluation
problem, together with a pointwise correctness theorem for an arbitrary path
assignment. The global acceptance condition is then obtained by universal
quantification over path assignments, as a semantic specification rather than
as a more efficient classical decision procedure (Section~\ref{sec:Boolean}).

\item A Lean formalization of the circuit evaluator and its correctness
theorems, together with structural \textbf{DLDS} validity predicates and a
machine-checked bridge for the uncompressed simple-tree fragment. In that
fragment, a valid simple-tree \textbf{DLDS} satisfying the route and discharge
conditions formalized in Lean yields genuine circuit acceptance of its
extracted path. Extending this bridge to the full compressed case, including
the recursive Flow condition, is left for future work (Section~\ref{sec:lean}
and Section~\ref{future}).
\end{enumerate}

\textbf{Organization.} Section~\ref{DLDS} recalls Natural Deduction for $\mathsf{M}_{\supset}$ and the \textbf{DLDS} compression framework. Section~\ref{sec:Boolean} presents the circuit construction and acceptance semantics. Section~\ref{sec:lean} describes the Lean formalization and correctness theorems. Section~\ref{related} discusses related work, and Section~\ref{future} outlines quantum verification as future work.

\section{Background}
In purely implicational minimal logic, denoted $M_{\supset}$, implication is the
only logical connective. Accordingly, the Natural Deduction system for
$M_{\supset}$ consists of exactly two inference rules:
  \begin{center}
  \bottomAlignProof
  \AXC{$[A]$}\noLine
  \UIC{$\vdots$}\noLine
  \UIC{$B$}\Rlbl{$\imply$-Introduction}
  \UIC{$A\imply{B}$}
  \DP
  \qquad
  \bottomAlignProof
  \AXC{$A$}
  \AXC{$A\imply{B}$}\Rlbl{$\imply$-Elimination}
  \BIC{$B$}
  \DP
  \end{center}
The implication introduction rule $\supset$I discharges all open assumptions of
$A$, while implication elimination $\supset$E corresponds to modus ponens.  An \textbf{ND} proof is a derivation in which no assumptions remain undischarged: every hypothesis introduced must eventually be closed by an implication introduction rule. 
Following \cite{haeusler2025horizontal}, such proofs admit a compressed representation as a \textbf{DLDS}. A \textbf{DLDS} is a directed acyclic graph (DAG) obtained from a tree-like \textbf{ND} proof by the \textbf{HC} algorithm: when the same formula appears multiple times at the same derivation level, these occurrences are merged into a single node. Each node represents a formula occurrence, and edges represent inference steps. We work exclusively with \emph{normal} \textbf{ND} proofs adopting
\emph{greedy discharge}~\cite{haeusler2025horizontal}: each $\supset$I
simultaneously discharges all open occurrences of its hypothesis. By strong
normalization for $M_\supset$~\cite{prawitz2006natural}, every provable formula
admits a normal proof. Moreover, normal proofs in $M_\supset$ satisfy the
\emph{subformula property}: every formula occurrence in a normal proof of
$\alpha$ is a subformula of $\alpha$. Greedy discharge is no restriction: every
proof can be rewritten into one in which every $\supset$I application is greedy,
preserving both the conclusion and normality~\cite[Lemma~3, Cor.~8]{haeusler2025horizontal}.
We may therefore assume throughout that proofs are normal and use greedy
discharge. The subformula property gives a finite, bounded universe of formulas,
while greedy discharge allows dependency information to be represented as sets of
active assumptions over that universe.
To avoid explicit discharge edges, each deduction edge carries a dependency
bitstring, a sequence of bits encoding which assumptions remain active at that
point in the derivation, where bit $i$ indicates whether the $i$-th subformula is
active. Throughout this paper, we refer to these bitstrings as \emph{dependency
sets}.

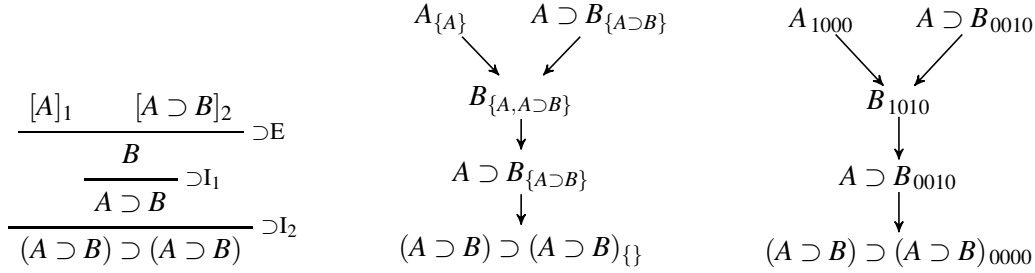
\begin{figure}[htbp]
  \centering
  \bottomAlignProof
\AXC{\strut$[A]_1$}
\AXC{\strut$[A \imply B]_2$}
\Rlbl{{$\imply$E}}
\BIC{\strut$B$}
\Rlbl{{$\imply$I$_1$}}
\UIC{\strut$A \imply B$}
\Rlbl{{$\imply$I$_2$}}
\UIC{\strut$(A \imply B) \imply (A \imply B)$}
\DisplayProof
\hfil%
\adjustbox{valign=b}{%
\begin{tikzpicture}[->,>=stealth',shorten >=1pt,node distance=2em, semithick,
  every node/.style={draw=none,anchor=base,inner sep=1pt,outer sep=0pt}]
  \tikzmath{let \y=2.83em;}
  \node (a) {\strut$A_{\{A\}}$};
  \node (imp) [right=of a] {\strut$A \imply B_{\{A \imply B\}}$};
  \node (b) at ($(a.base)!.5!(imp.base)-(0,\y)$){\strut$B_{\{A,\,A \imply B\}}$};
  \node (final) at ($(b)-(0,\y)$) {\strut$A \imply B_{\{A \imply B\}}$};
  \node (veryfinal) at ($(final)-(0,\y)$) {\strut$(A \imply B) \imply (A \imply B)_{\{\}}$};
  \path (a) edge (b);
  \path (imp) edge (b);
  \path (b) edge (final);
  \path (final) edge (veryfinal);
\end{tikzpicture}}%
\hfil
\adjustbox{valign=b}{%
\begin{tikzpicture}[->,>=stealth',shorten >=1pt,node distance=2em, semithick,
  every node/.style={draw=none,anchor=base,inner sep=1pt,outer sep=0pt}]
  \tikzmath{let \y=2.83em;}
  \node (a) {$A_{\,1000}$};
  \node (imp) [right=of a] {$A \imply B_{\,0010}$};
  \node (b) at ($(a.base)!.5!(imp.base)-(0,\y)$){$B_{\,1010}$};
  \node (final) at ($(b)-(0,\y)$) {$A \imply B_{\,0010}$};
  \node (veryfinal) at ($(final)-(0,\y)$) {$(A \imply B) \imply (A \imply B)_{\,0000}$};
  \path (a) edge (b);
  \path (imp) edge (b);
  \path (b) edge (final);
  \path (final) edge (veryfinal);
\end{tikzpicture}}%


  \caption{Three representations of the same minimal implicational proof: 
  (left) \textbf{ND} derivation, 
  (middle) dependency sets, 
  (right) bitstring encoding.}
  \label{fig:implication-dep-sets}
\end{figure}

\begin{definition}\label{def:bitstring}
Let $\alpha$ be an implicational formula. Denote by $Sub(\alpha)$ the set of all subformulas of $\alpha$, and let $\mathcal{O}(\alpha) = \{\beta_0, \beta_1, \ldots, \beta_k\}$ be a linear ordering of $Sub(\alpha)$. A \emph{bitstring} over $\mathcal{O}(\alpha)$ is a sequence $b_0 b_1 \ldots b_k$ with each $b_i \in \{0,1\}$ for $i=0,\ldots,k$. There is a bijection between bitstrings on $\mathcal{O}(\alpha)$ and subsets of $Sub(\alpha)$, given by $Set(b_0 b_1 \ldots b_k) \;=\; \{\beta_i \mid b_i = 1\}.$ The bitstring associated with $\mathcal{O}(\alpha)$ is used to represent which subformulas are selected, allowing us in particular to register discharged assumptions and the dependencies propagated through the proof. 
\end{definition}
This encoding scheme is illustrated in Fig.~\ref{fig:implication-dep-sets}, which 
follows the ordering $A \prec B \prec A \imply B \prec (A \imply B) \imply (A \imply B)$, 
showing the \textbf{ND} derivation, the dependency sets, and their bitstring encodings. 
Dependency sets evolve deterministically under rule application.
For implication elimination, the dependency set at the conclusion is the union of 
the dependency sets of both premises.
For implication introduction discharging a hypothesis $\varphi$, the formula $\varphi$ 
is removed from the dependency set.
Consequently, a derivation is closed if and only if the dependency set at the 
conclusion is empty.

We use two running examples. The first is $(A \imply B) \imply (A \imply B)$. Because this derivation has no repeated formula labels at the same level, it does not
admit horizontal compression. However, its small size allows us to fully present the circuit construction more clearly.
The second, which we will introduce later, is a larger tautology that contains same-level repetitions and thus yields a non-trivial compressed \textbf{DLDS} under the \textbf{HC} algorithm.

Intuitively, a \textbf{DLDS} is obtained by merging nodes with identical formulas 
that occur at the same level of an \textbf{ND} derivation, producing a DAG. Each 
node retains its formula label, while each deduction edge carries a dependency 
bitstring tracking which assumptions remain active. Additional \emph{ancestor edges} 
record the provenance introduced by node merges, preserving the logical reading of 
the original derivation. In this way, redundancy is eliminated while retaining 
enough information to verify each inference step. 
The \textbf{HC} algorithm is based on a finite collection of local
compression rules. Each rule identifies a specific pattern in a \textbf{DLDS},
typically two nodes labelled with the same formula at the same level, and
specifies how to collapse them while updating deduction edges, ancestor edges,
and dependency labels. The complete family consists of 28 such rules
~\cite{haeusler2025horizontal}. These rules are not used here as an unrestricted
abstract rewriting system. Rather, they are used inside the deterministic
\textbf{HC} algorithm: rules are applied only in the prescribed algorithmic
position, proceeding level by level from the conclusion side towards the
assumptions and, within each level, according to the fixed left-to-right order.

Thus, $\HC(\Pi)$ denotes the output of this fixed strategy applied to a
tree-like greedy derivation $\Pi$. We do not assume, and do not need, a
confluence theorem for arbitrary orders of rule application. For the purposes of
this paper, the relevant properties are those of the algorithmic strategy:
applications in algorithmic position preserve \textbf{DLDS} validity, the
procedure terminates, and the resulting compressed structure has no level
containing two nodes with the same formula label~\cite{haeusler2025horizontal}.
Consequently, the polynomial size bound used below does not rely on confluence of
the rewrite system, but on the bounded formula universe and on the level-wise
collapse performed by the \textbf{HC} algorithm.

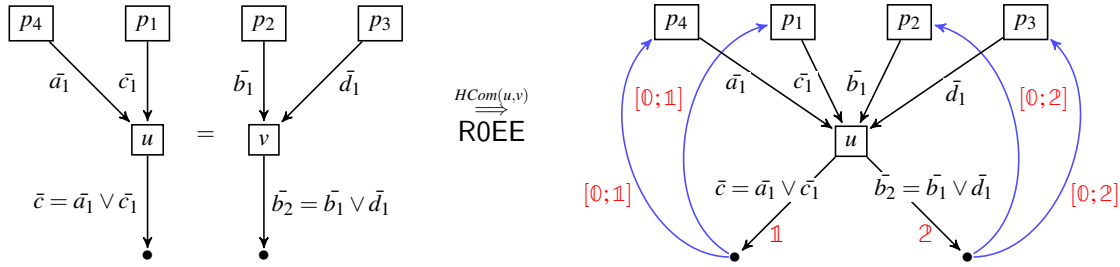
\begin{figure}[htbp]
  \centering
  \adjustbox{valign=c}{%
\begin{tikzpicture}
  [->,>=stealth',shorten >=1pt,semithick,node distance=4em,
every node/.style={draw=black,anchor=base,font=\footnotesize},
every edge quotes/.style={draw=none,font=\footnotesize,
  fill=white!50,inner sep=1pt,outer sep=2pt},
bul/.style={draw=none,inner sep=0pt,outer sep=0pt},
ord/.style={draw=none,right,red,font=\small},
ans/.style={draw=none,red,font=\footnotesize}]
\node (q1) {$p_1$};
\node (qn1) [left of=q1] {$p_4$};
\node (q2) [below of=q1] {$u$};
\node[bul] (q3) [below of=q2] {$\bullet$};
\node (q6) [right of=q1] {$p_2$};
\node (q7) [below of=q6] {$v$};
\node[bul] (q8) [below of=q7] {$\bullet$};
\node (q11) [right of=q6] {$p_3$};
\node[draw=none] (eq) at ($(q2)!.5!(q7)$) {$=$};
\path (q1) edge["$\bar{c_1}$" left] (q2)
(qn1) edge["$\bar{a_1}$" {left,xshift=-4pt}] (q2)
(q2) edge["$\bar{c}=\bar{a_1}\lor\bar{c_1}$" left] (q3)
(q6) edge["$\bar{b_1}$" left] (q7)
(q7) edge["$\bar{b_2}=\bar{b_1}\lor\bar{d_1}$" right] (q8)
(q11) edge["$\bar{d_1}$" {right,xshift=4pt}] (q7);
\end{tikzpicture}}
\hfil
$\stackrel{\stackrel{HCom(u,v)}{\Longrightarrow}}{\mathsf{R0EE}}$
\hfil
\adjustbox{valign=c}{%
\begin{tikzpicture}
  [->,>=stealth',shorten >=1pt,semithick,node distance=4em,
every node/.style={draw=black,anchor=base,font=\footnotesize},
every edge quotes/.style={draw=none,font=\footnotesize,
  fill=white!50,inner sep=1pt,outer sep=2pt},
bul/.style={draw=none,inner sep=0pt,outer sep=0pt},
ord/.style={draw=none,right,red,font=\small},
ans/.style={draw=none,red,font=\footnotesize}]
\node(q1) {$p_1$};
\node(qn1) [left of=q1] {$p_4$};
\node(q6) [right of=q1] {$p_2$};
\coordinate(q1q6) at ($(q1.base)!.5!(q6.base)$);
\node(q2) [below of=q1q6] {$u$};
\node[bul](q3) [below of=q2,xshift=-4em] {$\bullet$};
\node[bul](q8) [below of=q2,xshift=4em] {$\bullet$};
\node(q11) [right of=q6] {$p_3$};

\path (q8) [blue!70,bend right=70] edge node[ans,right,pos=.4] {$[\R0;\R2]$} (q11)
(q8) [blue!70,bend right=70] edge node[ans,right,pos=.6] {$[\R0;\R2]$} (q6)
(q3) [blue!70,bend left=70] edge node[ans,left,pos=.4] {$[\R0;\R1]$} (qn1)
(q3) [blue!70,bend left=70] edge node[ans,left,pos=.6] {$[\R0;\R1]$} (q1);

\path (q1) edge["$\bar{c_1}$" left] (q2)
(qn1) edge["$\bar{a_1}$" {left,xshift=-4pt}] (q2)
(q2) edge["$\bar{c}=\bar{a_1}\lor\bar{c_1}$" {pos=.4,xshift=-1em}] node[ord,pos=.8,right]{\R1} (q3)
(q6) edge ["$\bar{b_1}$" {left,xshift=-1pt}] (q2)
(q2) edge ["$\bar{b_2}=\bar{b_1}\lor\bar{d_1}$" {pos=.4,xshift=1em}] node[ord,pos=.8,left]{\R2} (q8)
(q11) edge ["$\bar{d_1}$" {right,xshift=6pt,pos=.6}] (q2);
\end{tikzpicture}}


  \caption{Before (left) and after (right) the collapse of nodes $u$ and
  $v$. Original deduction edges have default color $\R0$; non-zero
  colors introduced by the collapse are shown as red ordinals.}
  \label{fig:R0EE}
\end{figure}

Fig.~\ref{fig:R0EE} illustrates the role played by \emph{ancestor
edges} in \textbf{DLDS} structures. When the \textbf{HC} rule
\textsf{R0EE} collapses two nodes $u$ and $v$ labelled with the same
formula, the resulting structure ceases to be a tree and becomes a DAG.
This horizontal collapse is essential for compression, but it removes
the unique tree-like reading of the original \textbf{ND} derivation.
Ancestor edges are introduced precisely to record the provenance lost by
the collapse: they indicate how the collapsed node should be read along
the branches of the original derivation.

In these diagrams, every deduction edge carries two pieces of
information. Its \emph{color} determines membership in one of the sets
$E_D^i$, while its dependency label is the bitstring assigned by $L$. In
the uncompressed tree-like derivation all deduction edges have color
$\R0$; this color is omitted from the left-hand side of
Fig.~\ref{fig:R0EE} to reduce clutter. Thus, the edges from the premises
$p_i$ to $u$ or $v$ are color-$\R0$ deduction edges, whereas labels such
as $\bar a_1,\bar b_1,\bar c_1,\bar d_1$ are dependency bitstrings. The
non-zero red ordinals introduced on the right-hand side, such as $\R1$
and $\R2$, are new deduction-edge colors used to distinguish the
branches created by the collapse. Ancestor-edge labels such as
$[\R0;\R1]$ and $[\R0;\R2]$ are therefore paths of colors.

In Fig.~\ref{fig:R0EE} (right), the blue ancestor edges are labelled by
paths such as $[\R0;\R1]$ and $[\R0;\R2]$. A label $[\R0;\R1]$ means:
follow a deduction edge of color $\R0$ and then one of color $\R1$.
Thus, although multiple formula occurrences have been merged into one
node, their corresponding original branches remain distinct at the path level.
Bitstring expressions use $\vee$ for bitwise OR: for example, if two
branches depend on $\bar b_1$ and $\bar d_1$, their merge carries
$\bar b_1 \vee \bar d_1$.

Without ancestor edges, the collapsed structure would lose the
information needed to recover the original tree-like readings, permitting
unsound mixing of premises across distinct derivation branches. Ancestor
edges restore exactly this provenance, ensuring that dependency sets
propagate along the same logical branches as in the original derivation.
\begin{definition}[Dag-Like Derivability Structure]\label{DLDS}
Fix a set \(\Gamma\) of \(M_{\imply}\)-formulas and a linear order
\(\mathcal{O}_\Gamma\) on \(\Gamma\). Let
\[
\mathcal{O}_\Gamma^{0}:=\mathcal{O}_\Gamma\cup\{\R0,\lambda\},
\]
where \(\R0\) and \(\lambda\) are fresh symbols. The symbol \(\R0\) is the
default colour of deduction edges in the original tree-like derivation; non-zero
colours are introduced by horizontal collapse rules to distinguish branches
created by node merging. The symbol \(\lambda\) is reserved for deduction edges
whose dependency bitstring is not stored statically but is computed by the Flow
construction.

A \textbf{DLDS} is a tuple
\[
\mathcal{D}=
\langle V,r,l,(E_D^i)_{i\in\mathcal{O}_\Gamma^{0}},E_A,L,P\rangle
\]
where \(V\) is a nonempty set of nodes with root \(r\in V\) and label map
\(l:V\to\Gamma\); for each \(i\in\mathcal{O}_\Gamma^{0}\),
\(E_D^i\subseteq V\times V\) is the set of deduction edges of colour \(i\);
\(E_A\subseteq V\times V\) is the set of ancestor edges;
\[
L:\bigcup_{i\in\mathcal{O}_\Gamma^{0}}E_D^i
  \to \{0,1\}^{|\mathcal{O}_\Gamma|}
\]
labels deduction edges with dependency bitstrings; and
\[
P:E_A\to(\mathcal{O}_\Gamma^{0})^{*}
\]
assigns to each ancestor edge a finite colour path, its relative address.
\end{definition}

In a \textbf{DLDS}, the \emph{level} of a node measures its distance from the root along deduction edges. The root is at level~0. A node reachable from the root by a chain of $k$ deduction steps is at level~$k$. A \textbf{DLDS} is \emph{compressed} if, at each level, no two nodes share the same formula label. The example in Fig.~\ref{fig:implication-dep-sets} has no repetitions, making it compressed but, as we mentioned previously, it is a poor example for showcasing the \textbf{HC} algorithm. We therefore present a slightly more complex example. Fig.~\ref{fig:trimmed-proof} shows an \textbf{ND} proof of $(A_2\imply(A_3\imply A_4))\imply((A_1\imply(A_2\imply A_3))\imply((A_1\imply A_2)\imply(A_1\imply A_4)))$.

\begin{figure}[htbp]
  \input{figs/fig3}
  \caption{\textbf{ND} proof of $(A_2\imply(A_3\imply A_4))\imply((A_1\imply(A_2\imply A_3))\imply((A_1\imply A_2)\imply(A_1\imply A_4)))$.}
\label{fig:trimmed-proof}
\end{figure}

In Fig.~\ref{fig:trimmed-proof}, dependency sets are represented as bitstrings of length~13 (one bit per subformula), with the ordering:
\[
\begin{array}{l}
A_1 \prec A_2 \prec A_3 \prec A_4 \prec A_1 \imply A_2 \prec A_2 \imply A_3 \prec A_1 \imply A_4 
\prec A_3 \imply A_4  
\\ \prec A_1 \imply (A_2 \imply A_3) \prec A_2 \imply (A_3 \imply A_4) \prec (A_1\imply A_2)\imply(A_1\imply A_4) 
\\ \prec (A_1 \imply (A_2 \imply A_3)) \imply ((A_1\imply A_2)\imply(A_1\imply A_4)) 
\\ \prec (A_2\imply(A_3\imply A_4))\imply((A_1\imply(A_2\imply A_3))\imply((A_1\imply A_2)\imply(A_1\imply A_4)))
\end{array}
\]

\begin{figure}[htbp]
  \centering
  \let\Ground\bullet
\begin{tikzpicture}
[->,>=stealth',shorten >=1pt,semithick,node distance=2em,
every node/.style={draw=black,anchor=base,font=\footnotesize},
every edge quotes/.style={draw=none,font=\scriptsize,fill=white!50,inner sep=1pt,outer sep=2pt}]

\node(qp0){$A_1$};
\node[right=of qp0](qpUM){$A_1\imply A_2$};
\node[right=of qpUM](qp2){$A_1$};
\node[right=of qp2](qp3){$A_1\imply(A_2\imply A_3)$};
\node[right=of qp3](qp4){$A_1$};
\node[right=of qp4](qp5){$A_1\imply A_2$};

\tikzmath{let \y=-3.6em;}
\node(q0UM) at ($(qp0.base)!.5!(qpUM.base)+(0,\y)$){$A_2$};
\node(q23) at ($(qp2.base)!.5!(qp3.base)+(0,\y)$) {$A_2\imply A_3$};
\node(q45) at ($(qp4.base)!.5!(qp5.base)+(0,\y)$){$A_2$};
\node[right=of q45](qp6){$A_2\imply(A_3\imply A_4)$};

\node(q0UM23) at ($(q0UM.base)!.5!(q23.base)+(0,\y)$){$A_3$};
\node(q456) at ($(q45.base)!.5!(qp6.base)+(0,\y)$){$A_3\imply A_4$};

\node(q0UM2345) at ($(q0UM23.base)!.5!(q456.base)+(0,\y)$){$A_4$};

\node[draw=none,inner sep=0pt,outer sep=0pt](qFINAL) at ($(q0UM2345)+(0,\y)$){\strut$\cdots$};

\node(qFINAL1) at ($(qFINAL.south)+(0,\y/2)$){$(A_2\imply(A_3\imply A_4))\imply(A_1\imply(A_2\imply A_3))\imply(A_1\imply A_2)\imply(A_1\imply A_4)$};

\node[draw=none,inner sep=0pt,outer sep=0pt](qGROUND) at ($(qFINAL1)+(0,\y)$) {$\Ground$};

\path[->](qp0) edge["100000000000"{pos=.4,xshift=\y/3}] (q0UM);
\path[->](qpUM) edge["000010000000"{pos=.6,xshift=-\y/3}] (q0UM);

\path[->](qp2) edge["100000000000"{pos=.4,xshift=\y/4}] (q23);
\path[->](qp3) edge["000000001000"{pos=.6,xshift=-\y/4}] (q23);

\path[->](qp4) edge["100000000000"{pos=.4,xshift=\y/3}] (q45);
\path[->](qp5) edge["000010000000"{pos=.7,xshift=-\y/4}] (q45);

\path[->](q0UM) edge["100010000000"{pos=.4}] (q0UM23);
\path[->](q23) edge["100000001000"{pos=.6}] (q0UM23);

\path[->](q45) edge["100010000000"{pos=.4,xshift=\y/4}] (q456);
\path[->](qp6) edge["000000000100"{pos=.6,xshift=-\y/4}] (q456);

\path[->](q0UM23) edge["100010001000"] (q0UM2345);
\path[->](q456) edge["100010000100"] (q0UM2345);

\path[->](q0UM2345) edge["100010001100"] (qFINAL);
\path[->](qFINAL) edge (qFINAL1);
\path[->](qFINAL1)edge["000000000000"] (qGROUND);
\end{tikzpicture}


  \caption{Uncompressed \textbf{DLDS} of the \textbf{ND} proof in Fig.~\ref{fig:trimmed-proof}. Each node is labeled with its formula and dependency set. Note that $A_2$ appears twice at the same level, showing that the derivation has not yet been horizontally compressed.}
  \label{fig:graph-derivation1}
\end{figure}
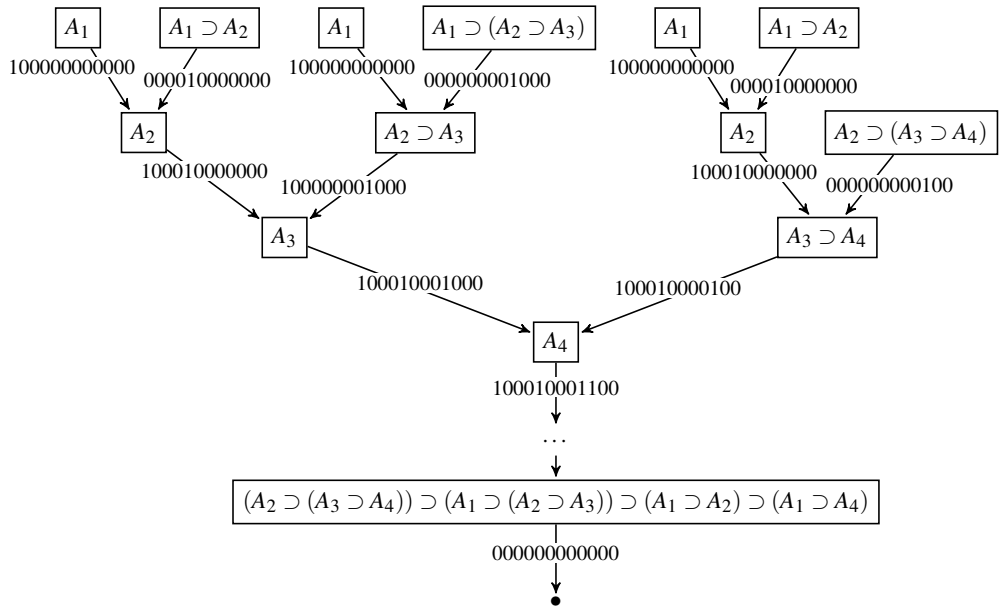

Fig.~\ref{fig:graph-derivation1} shows the upper part of the \textbf{DLDS} representation of the proof in Fig.~\ref{fig:trimmed-proof}. This is an uncompressed \textbf{DLDS}, since we have repetitions of the same formula at the same level, namely $A_2$ at level~7. The ``$\cdots$'' in Fig.~\ref{fig:graph-derivation1} abbreviates the entire sequence of consecutive implication introduction applications that discharge the remaining open assumptions and lead to the conclusion formula. These introduction steps are standard and structurally uninformative for the purposes of illustrating horizontal compression, so they are visually collapsed into a single arrow to improve readability while preserving the logical content of the derivation. Observe that the dependency set at the conclusion node contains the empty set of dependencies; hence, the conclusion is a tautology.

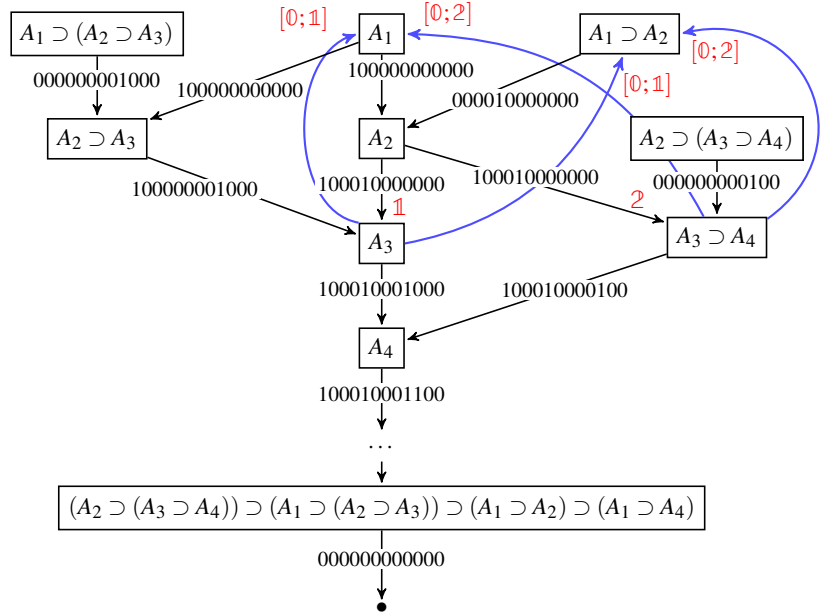
\begin{figure}[htbp]
  \centering
  \let\Ground\bullet
\begin{tikzpicture}
[->,>=stealth',shorten >=1pt,semithick,node distance=2em,
every node/.style={draw=black,anchor=base,font=\footnotesize},
every edge quotes/.style={draw=none,font=\scriptsize,fill=white!50,inner sep=1pt,outer sep=2pt},
ord/.style={draw=none,right,red,font=\small},
ans/.style={draw=none,red,font=\footnotesize}]

\begin{scope}[node distance=6em]
\node(A1){$A_1$};
\node[right=of A1](A1impA2){$A_1 \imply A_2$};
\node[left=of A1](A1impA2impA3){$A_1 \imply (A_2 \imply A_3)$};
\end{scope}

\tikzmath{let \y=-3.6em;}
\node(A2) at ($(A1.base)+(0,\y)$){$A_2$};
\node(A2impA3) at ($(A1impA2impA3.base)+(0,\y)$){$A_2 \imply A_3$};
\node(A2impA3A4phantom) at ($(A1impA2.base)+(3em,\y)$){\phantom{$A_2 \imply (A_3 \imply A_4)$}};

\node(A3) at ($(A2.base)+(0,\y)$){$A_3$};
\node(A3impA4) at ($(A2impA3A4phantom)+(0,\y)$){$A_3 \imply A_4$};

\node(A4) at ($(A3.base)+(0,\y)$){$A_4$};

\node[draw=none,inner sep=0pt,outer sep=0pt](qFINAL) at ($(A4)+(0,\y)$){\strut$\cdots$};

\node(qFINAL1) at ($(qFINAL.south)+(0,\y/2)$){$(A_2\imply(A_3\imply A_4))\imply(A_1\imply(A_2\imply A_3))\imply(A_1\imply A_2)\imply(A_1\imply A_4)$};

\node[draw=none,inner sep=0pt,outer sep=0pt](qGROUND) at ($(qFINAL1)+(0,\y)$) {$\Ground$};

\path[blue!70,thick,bend left=80]
  (A3.north west) edge node[ans,pos=.9,above left]{$[\R{0};\R{1}]$} (A1.west);
\path[blue!70,thick,bend right=30]
  (A3.east) edge node[ans,pos=.9,right]{$[\R{0};\R{1}]$} (A1impA2);
\path[blue!70,thick,bend right=80,looseness=1.2,min distance=4.4em,overlay]
  (A3impA4) edge node[ans,pos=.9,below]{$[\R{0};\R{2}]$} (A1impA2.east);
\path[blue!70,thick,bend right=30, looseness=1.1]
  ($(A3impA4.north)-(5pt,0)$) edge node[ans,pos=.9,above]{$[\R{0};\R{2}]$} (A1.east);

\node[fill=white](A2impA3A4) at ($(A1impA2.base)+(3em,\y)$){$A_2 \imply (A_3 \imply A_4)$};

\path(A1) edge["100000000000" {pos=.35,xshift=1em}](A2);
\path(A1impA2) edge["000010000000" {pos=.7,xshift=2em}](A2);
\path(A1) edge["100000000000" {pos=.7,xshift=1em}](A2impA3);
\path(A1impA2impA3) edge["000000001000"](A2impA3);
\path(A2) edge["100010000000"]
          node[ord,near end]{$\R{1}$}(A3);
\path(A2impA3) edge["100000001000" {xshift=-2em}] (A3);

\path(A2impA3A4) edge ["000000000100"](A3impA4);
\path(A2) edge["100010000000"]
         node[ord,pos=.8,xshift=2pt,yshift=3pt]{$\R{2}$}(A3impA4);

\path(A3) edge["100010001000"] (A4);
\path(A3impA4) edge["100010000100" xshift=1em] (A4);

\path[->](A4) edge["100010001100"] (qFINAL);
\path[->](qFINAL) edge (qFINAL1);
\path[->](qFINAL1) edge["000000000000"] (qGROUND);

\end{tikzpicture}


  \caption{Compressed \textbf{DLDS} obtained by applying the \textbf{HC} algorithm to Fig.~\ref{fig:graph-derivation1}. Identical formulas at the same level are collapsed into a single node, with new ancestor edges and labels added to preserve logical correctness.}
  \label{fig:dlds-compressed}
\end{figure}

Fig.~\ref{fig:dlds-compressed} shows the result of applying \textbf{HC}~\cite{haeusler2025horizontal} to the
\textbf{DLDS} shown in Fig.~\ref{fig:graph-derivation1}. The \textbf{HC} algorithm collapses, from left-to-right, the first pair of nodes with the same formula label. New ancestor edges and edge labels are added to preserve the logical reading of the \textbf{DLDS}, while one of the two nodes is deleted. In this example, when the two nodes labeled $A_2$ at level~7 are collapsed, the algorithm creates two ancestor edges and places the labels $\R1$ and $\R2$ on the deduction edges $A_2$ to $A_3$ and $A_2$ to $(A_3 \imply A_4)$, respectively, to preserve the derivability relation present before the collapse. Initially, two ancestor edges are created: one from $A_3$ to $A_1 \supset A_2$, labeled with the path $[\R0, \R1]$, and another from $A_3$ to $A_1$, also labeled $[\R0, \R1]$. These edges record that $A_3$ is the conclusion of a path starting at $A_1 \supset A_2$ and the conclusion of a path starting at $A_1$. To verify that
a \textbf{DLDS} is valid, one must locally check correct rule applications (including updates to dependency sets). 

The \textbf{HC} compression algorithm collapses nodes with the same formula label.
A node without incoming deduction edges is a \emph{top-node}, corresponding to an
open hypothesis occurrence in the original tree-like derivation. Some compression
rules collapse a top-node with another node, or collapse two nodes at least one of
which represents a hypothesis occurrence. In these cases, the right-hand side of
the rule carries an \(h\)-marking, indicating that the resulting collapsed node
must still be treated as hypothesis-originating. Thus, the marking is not an
additional formula label or edge label; it is a bookkeeping flag used by the
\textbf{HC} rules to preserve the hypothesis/top-node status needed when defining
ancestor-guided paths and the ancestor-simplicity condition below. Equivalently,
in the auxiliary notions below, a node may count as top either because it has no
incoming deduction edge or because it carries this hypothesis marker. We also use the following auxiliary notions. Let
\[
E_D := \bigcup_{i\in\mathcal{O}_{\Gamma}^{0}} E_D^i
\]
be the set of all deduction edges, forgetting colors. A \emph{deductive path}
from \(v\) to \(w\) is a finite non-empty sequence of vertices
$
v=u_0,u_1,\ldots,u_k=w, (k\geq 1)
$
such that, for every \(0\leq j<k\), the edge
\(\langle u_j,u_{j+1}\rangle\) belongs to \(E_D\). Ancestor edges are not
steps of a deductive path; they only provide provenance information used to
guide the reading of such paths. For \(w\in V\), define
$
Pre(w)=\{\,v\in V \mid \text{there exists a deductive path from }v
\text{ to }w\,\}.
$
For a non-empty color path \(p=o_1;\ldots;o_n\), let
\(\mathit{head}(p)=o_1\) and \(\mathit{tail}(p)=o_2;\ldots;o_n\). A
\emph{residual color path} is an element of
\((\mathcal{O}_{\Gamma}^{0})^{*}\). In a pair \((\vec b,p)\), the path
\(p\) records the remaining sequence of colors that must be followed from the
current node toward the target node.

For a formula \(\alpha\in\Gamma\), let \(\vec b_{\alpha}\in
\{0,1\}^{|\mathcal{O}_\Gamma|}\) denote the characteristic bitstring of the
singleton set \(\{\alpha\}\) with respect to the fixed order
\(\mathcal{O}_\Gamma\). Thus \(\vec b_{l(v)}\) is the dependency bitstring
corresponding to the formula labelling \(v\).

\begin{definition}[Flow]\label{def:flow}
Let
$
\mathcal{D}=
\langle V,(E_D^i)_{i\in\mathcal{O}_\Gamma^{0}},E_A,r,l,L,P\rangle
$
be a \textbf{DLDS}, and let \(w\in V\). The function
\[
\operatorname{Flow}(\mathcal{D},w) : Pre(w) \to
  \mathcal{P}\bigl(\{0,1\}^{|\mathcal{O}_\Gamma|}
  \times(\mathcal{O}_\Gamma^{0})^{*}\bigr)
\]
assigns to each \(v\in Pre(w)\) a set of pairs \((\vec b,p)\), where
\(\vec b\) is a dependency bitstring and \(p\) is a residual color path. Each
pair records one ancestor-guided deduction route from \(v\) to \(w\), together
with the dependencies carried along that route.

The main recursive clauses are as follows. If \(v\) is a top-node, then Flow
starts with the dependency bitstring \(\vec b_{l(v)}\). If \(v\) has no incoming
ancestor edge, this contributes \((\vec b_{l(v)},\varepsilon)\); for each
incoming ancestor edge \(e=\langle v',v\rangle\), it contributes
\((\vec b_{l(v)},P(e))\).

At an implication-elimination node with premises \(v_1\) and \(v_2\), compatible
premise pairs
\[
(\vec b_1,[o_1\mid p])\in \operatorname{Flow}(\mathcal{D},w)(v_1),
\qquad
(\vec b_2,[o_2\mid p])\in \operatorname{Flow}(\mathcal{D},w)(v_2)
\]
combine to produce
$
(\vec b_1\vee \vec b_2,p),
$
where \(\vee\) is bitwise OR. At an implication-introduction node discharging formula \(\alpha\), a premise
pair
$
(\vec b',[o'\mid p])
$
produces
$
(\vec b'\wedge \neg \vec b_{\alpha},p),
$
removing the discharged assumption from the dependency bitstring. The complete recursive definition, including the cases involving collapsed
nodes and \(\lambda\)-labelled edges, is given in
\cite[Def.~22]{haeusler2025horizontal}.
\end{definition}
We now define a valid \textbf{DLDS}.

\begin{definition}[Valid DLDS]\label{def:ValidDLDS}
A structure 
$
\mathcal{D} = \langle V , (E_{D}^{i})_{i \in \mathcal{O}_{\Gamma}^{0}}, E_{A}, r, l, L, P \rangle
$
is a valid \textbf{DLDS} if the following conditions hold:
\begin{description}
  \item[Color-Acyclicity.] For each $i \in \mathcal{O}_{\Gamma}^{0}$, the deduction edge set $E_D^i$ is acyclic.
  \item[Leveled-Colored.] The rooted sub-DAG $\langle V , (E_{D}^{i})_{i\in\mathcal{O}_{\Gamma}^{0}}, r\rangle$ is leveled.
  \item[Ancestor-Edges.] For every ancestor edge $\langle v_1, v_2 \rangle \in E_A$, the level of $v_1$ is less than the level of $v_2$.
  \item[Ancestor-Backway-Information.] Each ancestor edge $\langle v_1, v_2 \rangle$ is labeled by $P(\langle v_1, v_2 \rangle)$ with the relative address of $v_1$ from $v_2$.
  \item[Simplicity.] The rooted deduction graph is simple: for any pair of nodes, there is at most one deduction edge between them in a given color class.
  \item[Non-Nested-Ancestor-Edges.] No ancestor edge is nested along the path defined by another.
 \item[Flow Condition.] For each node \(w\in V\), the flow
\(\operatorname{Flow}(\mathcal D,w)(v)\) is defined for every \(v\in Pre(w)\).
If this set is a singleton \(\{(\vec b,p)\}\), then \(v\) has exactly one
outgoing deduction edge \(\langle v,v'\rangle\), this edge has color
\(\mathit{head}(p)\), belongs to \(E_D^{\mathit{head}(p)}\), and has label
\(L(\langle v,v'\rangle)=\vec b\). If the flow is non-empty and not a
singleton, then for each color \(i\), the subset \(\Phi_i\) of flow elements
with head \(i\) determines the unique outgoing edge of color \(i\): if
\(\Phi_i\neq\emptyset\), such an edge exists and is labelled by \(\vec b\) when
\(\Phi_i=\{(\vec b,p)\}\), and by \(\lambda\) otherwise; if
\(\Phi_i=\emptyset\), no such edge exists.
\end{description}
\end{definition}

In Fig.~\ref{fig:dlds-compressed}, we show the totally compressed and valid \textbf{DLDS} produced by the \textbf{HC} algorithm when it is applied to the derivation in Fig.~\ref{fig:graph-derivation1}. 

Validity of a \textbf{DLDS} is defined semantically via the $\operatorname{Flow}$ 
construction. Flow tracks how dependency sets propagate through the compressed 
structure by following ancestor-guided paths, ensuring dependencies are maintained 
exactly as in the original tree-like derivation. This guarantees that node merging 
does not create unsound mixing of premises: each path through the DAG corresponds 
to a valid branch in the original proof. The function $\operatorname{Flow}$, defined recursively above, satisfies:
if $\operatorname{Flow}(\mathcal{D},w)(v)=\{(\vec b_1,p_1),\ldots,(\vec b_k,p_k)\}$, then there are
$k$ distinct deduction paths from $v$ to $w$, each with dependency vector $\vec b_i$
and residual path $p_i$. In~\cite{haeusler2025horizontal,RobinsonLean}, we proved (formalized in Lean) the following.

\begin{theorem}
Let $\Pi$ be any \textbf{DLDS} obtained by direct translation from \textbf{ND}.
After applying the \textbf{HC} algorithm, we obtain a valid and compressed \textbf{DLDS}.
\end{theorem}

A totally compressed valid \textbf{DLDS} is the normal form obtained after the exhaustive application of the \textbf{HC} algorithm to a tree-like \textbf{ND} derivation in minimal implicational logic. It is a finite, leveled directed acyclic graph in which, at each level, there is at most one node for each formula, so that no further horizontal compression rule applies. Deductive edges encode the local inferential structure, while ancestor edges, labeled by finite paths, record the vertical provenance lost by node collapsing, allowing the DAG to be read as a family of valid tree-like derivations. Thus, a totally compressed \textbf{DLDS} is a maximally shared, sound representation of a proof, preserving logical correctness while admitting polynomial-time verification~\cite{haeusler2025horizontal}.

\section{From DLDS to Boolean Circuit}\label{sec:Boolean}
The \textbf{HC} algorithm transforms an \textbf{ND} proof into a compressed
\textbf{DLDS}, where nodes represent formulas and edges represent inference
steps. In this section, we describe how the formula universe of a \textbf{DLDS}
can be used to build a Boolean circuit evaluator for candidate derivation paths.
Rather than encoding a single derivation directly, the circuit is constructed in
a uniform way so as to accommodate all admissible local rule applications. Each
grid cell is compiled into a local subcircuit that checks the corresponding
candidate inference and propagates dependency information. The overall structure
of the circuit is fixed, while path variables dynamically select which potential
wires are active at each level. This yields a uniform evaluation mechanism for
candidate \textbf{DLDS} paths without explicitly enumerating derivation paths.

The Boolean circuit is constructed as a uniform grid from the formula universe
associated with the \textbf{DLDS}. Let \(N\) be the number of distinct formula
labels. Each row represents one derivation level and contains one cell for each
formula label. Between consecutive levels we include a complete set of
\emph{potential} wires: every cell at level \(i\) is connected to every cell at
level \(i+1\). These wires are part of the circuit scaffold; they are not the
deduction edges or ancestor edges of the original \textbf{DLDS}. The complete
grid provides a fixed execution space in which every candidate local predecessor
pattern can be represented. A path assignment activates only selected potential
wires, and the local subcircuits then check whether the selected inputs form an
admissible rule instance, namely \(\supset I\), \(\supset E\), or repetition.

Thus, the grid should be understood as an over-approximation of the original
\textbf{DLDS}: it contains many candidate connections that do not correspond to
actual derivation steps. Such candidates are rejected by the local rule-checking
subcircuits. This construction is independent of the \emph{Ancestor-Edges}
property of Definition~\ref{def:ValidDLDS}, which concerns only the actual
ancestor edges \(E_A\) of a \textbf{DLDS}.

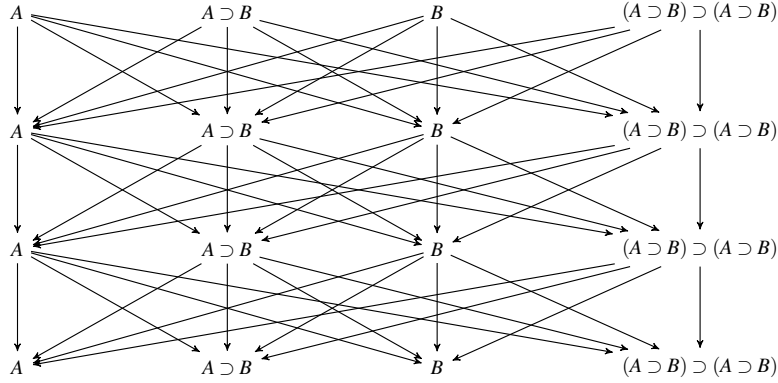
\begin{figure}[htbp]
\centering
\scalebox{0.7}{
\begin{tikzpicture}[scale=0.5,->,>=stealth',shorten >=1pt,node distance=1.7cm and 3.1cm, semithick]

  \node (A1) {$A$};
  \node (IMP1) [right=of A1] {$A \imply B$};
  \node (B1) [right=of IMP1] {$B$};
  \node (AB1) [right=of B1] {$(A \imply B) \imply (A \imply B)$};

  \node (A2) [below=of A1] {$A$};
  \node (IMP2) [right=of A2] {$A \imply B$};
  \node (B2) [right=of IMP2] {$B$};
  \node (AB2) [right=of B2] {$(A \imply B) \imply (A \imply B)$};

  \node (A3) [below=of A2] {$A$};
  \node (IMP3) [right=of A3] {$A \imply B$};
  \node (B3) [right=of IMP3] {$B$};
  \node (AB3) [right=of B3] {$(A \imply B) \imply (A \imply B)$};

  \node (A4) [below=of A3] {$A$};
  \node (IMP4) [right=of A4] {$A \imply B$};
  \node (B4) [right=of IMP4] {$B$};
  \node (AB4) [right=of B4] {$(A \imply B) \imply (A \imply B)$};

  \foreach \i in {A1,IMP1,B1,AB1} {
    \foreach \j in {A2,IMP2,B2,AB2} {
      \path (\i) edge (\j);
    }
  }
  \foreach \i in {A2,IMP2,B2,AB2} {
    \foreach \j in {A3,IMP3,B3,AB3} {
      \path (\i) edge (\j);
    }
  }
  \foreach \i in {A3,IMP3,B3,AB3} {
    \foreach \j in {A4,IMP4,B4,AB4} {
      \path (\i) edge (\j);
    }
  }

\end{tikzpicture}}
\caption{A fully connected \(4\times4\) circuit scaffold associated with the
formula universe of Fig.~\ref{fig:implication-dep-sets}. Each level contains one
node for each formula label, and potential wires connect all nodes across
successive levels.}
\label{fig:4x4grid}
\end{figure}

For illustration, the \textbf{DLDS} in
Fig.~\ref{fig:implication-dep-sets} has four distinct formulas: \(A\), \(B\),
\(A \imply B\), and \((A \imply B) \imply (A \imply B)\), so we use \(N = 4\).
Fig.~\ref{fig:4x4grid} depicts the resulting \(4\times4\) circuit grid, in which
each cell is implemented by a local subcircuit.

The grid topology is chosen for uniformity. It provides the same circuit
architecture for every formula universe of size \(N\), while the path assignment
selects which potential wires are active in a particular evaluation. If the grid
has \(N\) columns and at most \(N+1\) levels, then it has \(O(N^2)\) cells. Each
local subcircuit has polynomial size in \(N\), and therefore the whole grid has
polynomial size in \(N\). More precisely, under the gate accounting used here,
each local subcircuit has size \(O(N)\), yielding an \(O(N^3)\) circuit.

The exponentially large object is not the circuit itself, but the semantic space
of path assignments quantified over in the global acceptance condition defined
below.

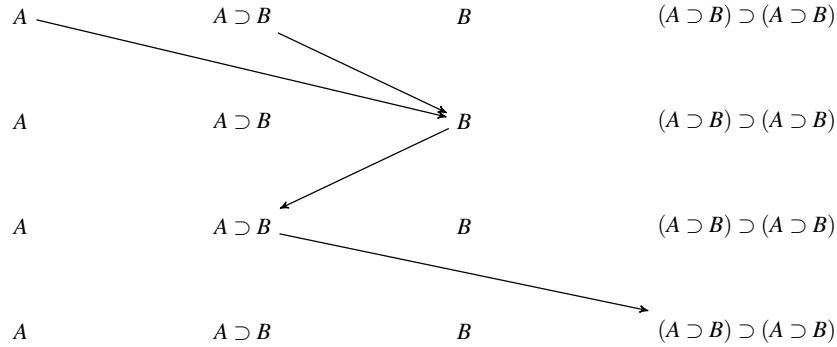
\begin{figure}[htbp]
\centering
\scalebox{0.8}{
\begin{tikzpicture}[scale=0.8,->,>=stealth',shorten >=1pt,node distance=1.2cm and 2.8cm, semithick]

  \node (A1) {$A$};
  \node (IMP1) [right=of A1] {$A \imply B$};
  \node (B1) [right=of IMP1] {$B$};
  \node (AB1) [right=of B1] {$(A \imply B) \imply (A \imply B)$};

  \node (A2) [below=of A1] {$A$};
  \node (IMP2) [right=of A2] {$A \imply B$};
  \node (B2) [right=of IMP2] {$B$};
  \node (AB2) [right=of B2] {$(A \imply B) \imply (A \imply B)$};

  \node (A3) [below=of A2] {$A$};
  \node (IMP3) [right=of A3] {$A \imply B$};
  \node (B3) [right=of IMP3] {$B$};
  \node (AB3) [right=of B3] {$(A \imply B) \imply (A \imply B)$};

  \node (A4) [below=of A3] {$A$};
  \node (IMP4) [right=of A4] {$A \imply B$};
  \node (B4) [right=of IMP4] {$B$};
  \node (AB4) [right=of B4] {$(A \imply B) \imply (A \imply B)$};

  \path (A1) edge (B2);
  \path (IMP1) edge (B2);

  \path (B2) edge (IMP3);

  \path (IMP3) edge (AB4);

\end{tikzpicture}}
    \caption{\(4\times4\) grid restricted by an assignment of the path variable.}
    \label{fig:small-example-extended}
\end{figure}

The circuit input is a \emph{path array}, which is not a single path but a tuple of
\(N\) independent traversals, one for each formula of the \textbf{DLDS}. Each subpath corresponds to a distinct formula, reflecting that dependency information is tracked separately for each formula and must be verified independently. We decompose it into \(N\) \emph{subpaths}, one per column of the \(N \times N\) grid. Intuitively, each subpath represents a candidate derivation starting from a given
top formula and progressing downward through the grid. Fig.~\ref{fig:small-example-extended} shows a subgrid corresponding to a particular path assignment: the path variables select which edges are active at each level, tracing a single derivation through the grid.

Formally, a subpath is a finite list
$
[(t_0,\ell_0),(t_1,\ell_1),\ldots,(t_{h-1},\ell_{h-1})],
$
where \(h\) is the number of transitions between consecutive levels of the
grid. The first component \(t_i\) selects the target column at level \(i+1\):
the value \(0\) means that the traversal stops, while \(t_i=j+1\) means that
the traversal moves to column \(j\). The second component \(\ell_i\) records
which input position of the target subcircuit is being used. This label is
needed because implication elimination has two premises, while implication
introduction and repetition have one. The full path assignment
\(P\in\mathsf{Path}_{N,h}\) consists of one such subpath for each formula
column. Taken together, these subpaths select a restricted subgraph of the
\(N\times N\) grid. In the figures and informal examples below, we often display only the target
components \(t_i\), suppressing the input labels \(\ell_i\).

Fig.~\ref{fig:4x4grid} depicts the full grid for a proof with four formulas. Each subcircuit in the top row receives a path variable, and together these variables determine a restricted subgraph. Consider the following assignment:
$\mathrm{Path}_1 = [3,2,4],  \mathrm{Path}_2 = [3,2,4],
\mathrm{Path}_3 = [0,0,0],
\mathrm{Path}_4 = [0,0,0].
$
Here, $\mathrm{Path}_1$ corresponds to the leftmost subcircuit in the top row. The first entry, “3” selects the third outgoing edge at level~1; the next entry, “2” selects the second outgoing edge at level~2; and so on. The assignment $\mathrm{Path}_3 = [0,0,0]$ indicates that the third subcircuit is inactive from the start (no edges are chosen). The fourth path is interpreted analogously. The effect of this assignment is that 
only a subset of the original edges remains active, yielding the restricted 
subgraph shown in Fig.~\ref{fig:small-example-extended}. The Boolean circuit constructed from the \textbf{DLDS} then evaluates whether
this induced subgraph is locally compatible with the rule instances represented
at each active subcircuit. A subcircuit is \emph{active} if it lies on a path selected by the path variables; each active subcircuit enforces one valid rule application and updates its dependency set. If all active rule applications are locally well formed and the final dependency
set output by the conclusion subcircuit is empty, the path assignment satisfies the pointwise acceptance condition defined below.

\textit{Acceptance semantics.}
Let \(h\) be the number of transitions between consecutive levels of the grid. We
write $
  \mathsf{Path}_{N,h}
  =
  \bigl((\{0,\ldots,N\}\times L)^{h}\bigr)^{N},
$
where \(L\) is the finite set of input labels used by the local subcircuits.
Thus a path assignment \(P\in\mathsf{Path}_{N,h}\) is an \(N\)-tuple $P = (\mathrm{Path}_1,\ldots,\mathrm{Path}_N)$,
where each \(\mathrm{Path}_i\) is a labeled sequence of \(h\) choices. In a
step \((t,\ell)\), the value \(t=0\) means that the corresponding path is
inactive at that step, while \(t=j\in\{1,\ldots,N\}\) selects the potential wire
going to column \(j\) in the next level. The label \(\ell\in L\) specifies the
input position of the target subcircuit.

For a fixed \textbf{DLDS} \(\mathcal D\), goal column \(c\), and path assignment
\(P\in\mathsf{Path}_{N,h}\), the circuit computes a pointwise acceptance bit
$
  \mathrm{acc}_{\mathcal D,c}(P)\in\{0,1\}.
$
This bit is defined in terms of three predicates:
\[
  \mathrm{Invalid}_{\mathcal D}(P),\qquad
  \mathrm{WellFormed}_{\mathcal D}(P),\qquad
  \mathrm{Discharged}_{\mathcal D,c}(P).
\]
Here \(\mathrm{Invalid}_{\mathcal D}(P)\) means that the active subgraph selected
by \(P\) contains a local rule conflict: at least one active subcircuit receives
selected inputs that do not determine exactly one admissible instance of
\(\supset I\), \(\supset E\), or repetition. Conversely,
\(\mathrm{WellFormed}_{\mathcal D}(P)\) means that every active local subcircuit
does determine exactly one such admissible rule instance. Finally,
\(\mathrm{Discharged}_{\mathcal D,c}(P)\) means that the dependency vector at the
goal column \(c\) is the all-zero bitstring.

The pointwise acceptance condition is
$
\mathrm{acc}_{\mathcal D,c}(P)=1
\quad\Longleftrightarrow\quad
\mathrm{Invalid}_{\mathcal D}(P)
\ \vee\
\bigl(
  \mathrm{WellFormed}_{\mathcal D}(P)
  \wedge
  \mathrm{Discharged}_{\mathcal D,c}(P)
\bigr).
$
Thus an ill-formed path assignment is assigned the acceptance value \(1\), so it
cannot by itself reject the universal check. This is not a claim that the
ill-formed assignment represents a derivation; it only means that such an
assignment is outside the locally well-formed candidate paths that can witness a
failure of discharge. A value \(0\) can arise only from a locally well-formed path
assignment whose final dependency vector is nonzero. The global acceptance condition is then the universal condition
$
  \mathrm{Accept}(\mathcal D,c)
  \quad:\Longleftrightarrow\quad
  \forall P\in\mathsf{Path}_{N,h},\
  \mathrm{acc}_{\mathcal D,c}(P)=1.
$
Equivalently, since \(\mathsf{Path}_{N,h}\) is finite,
$
  \mathrm{Accept}(\mathcal D,c)
  =
  \bigwedge_{P\in\mathsf{Path}_{N,h}}
  \mathrm{acc}_{\mathcal D,c}(P).
$

This conjunction ranges over all path assignments in the uniform grid, not only
over assignments that already correspond to genuine derivation paths. Because
\(|\mathsf{Path}_{N,h}|\) is exponential in the grid height and width,
\(\mathrm{Accept}(\mathcal D,c)\) is used here as a semantic specification, not
as a polynomial-time classical decision procedure. The Lean formalization proves
the pointwise statement for an arbitrary fixed \(P\); the global condition follows
by universal quantification over \(P\).

\begin{figure}[htbp]
    \centering
    \includegraphics[scale=0.9]{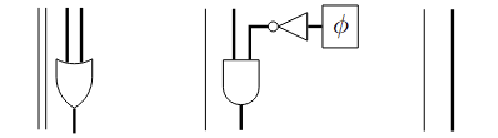}
    \caption{Circuit modules for the three inference rules: implication elimination
    ($\imply E$), implication introduction ($\imply I$), and repetition ($R$). Thick
    wires carry dependency sets; thin wires carry activation bits.}
    \label{fig:rules-box}
\end{figure}

We now describe how \textbf{DLDS} inference steps are compiled into circuit modules. Each 
subcircuit in the grid implements one of three operations: implication introduction 
($\supset$I), implication elimination ($\supset$E), or repetition (R). These modules, 
shown in Fig.~\ref{fig:rules-box}, share a common architecture: they accept dependency 
sets as input and produce an updated dependency set as output, following the 
propagation rules from Section~\ref{DLDS}.

In addition to the two standard \textbf{ND} rules, we include a \emph{repetition rule} 
following~\cite{GH2020}. This is a structural mechanism required for circuit construction. In the original \textbf{DLDS}, a hypothesis may first appear at any level depending on the derivation structure. However, in the uniform $N \times N$ grid, all formulas occupy the top row. To propagate a hypothesis from the top row down 
to the level where it is first used in the \textbf{DLDS}, we need an identity operation that 
passes the dependency set unchanged through intermediate levels. The repetition rule 
serves exactly this purpose. During circuit construction, repetition modules are inserted as needed to align the grid structure with the \textbf{DLDS} topology. Thus, while repetition is redundant from a logical standpoint, it is essential for maintaining a uniform circuit architecture.

Each subcircuit must determine which of the three inference operations ($\supset$I, 
$\supset$E, or R) applies for a given path assignment. Since multiple rules could 
potentially apply at any grid position, we use \emph{activation bits}, Boolean inputs 
that select which rule is actually active. The resulting dependency set is then routed 
through a Boolean \emph{selector} that implements the path variable's 
choice. Implication elimination requires two activation bits (one per premise); 
implication introduction and repetition each require one.

The three circuit modules are shown in Fig.~\ref{fig:rules-box}, where thick wires 
represent dependency sets and thin wires represent activation bits. Each module accepts 
activation bit(s) and incoming dependency set(s) from its premise(s), then computes an 
updated dependency set according to the rule's semantics. For implication elimination, 
the output is the bitwise OR of both incoming dependency sets: 
$\vec d^{\text{out}}=\vec d^{\text{in}}_1 \lor \vec d^{\text{in}}_2$. For implication 
introduction, the output is the bitwise AND of the incoming set with the negation of 
discharge mask $\chi_\phi$ (the one-hot bitstring for discharged hypothesis $\phi$): 
$\vec d^{\text{out}}=\vec d^{\text{in}} \land \neg \chi_\phi$. For repetition, 
$\vec d^{\text{out}}=\vec d^{\text{in}}$ (identity).

This architecture separates dependency propagation from structural invalidity.
When exactly one admissible rule instance is active, the subcircuit outputs the
dependency vector computed by that rule. When no rule is active, it outputs the
zero vector. When the selected active inputs fail to determine a unique admissible
rule instance, the path assignment is marked by
\(\mathrm{Invalid}_{\mathcal D}(P)\), as in the acceptance semantics above.

A node subcircuit contains one candidate gate for each admissible local rule
instance concluding the formula at that grid cell. Activation bits select the
rule instance, an XOR gate enforces uniqueness, and the selected dependency
vector is forwarded through the path selector. If selected inputs do not
determine exactly one admissible rule instance, the path assignment contributes
to \(\mathrm{Invalid}_{\mathcal D}(P)\). For example, at a cell labelled \(B\),
the inputs \(A\) and \(A\imply B\) activate an implication-elimination gate,
whereas an incoming wire that cannot participate in any admissible rule instance
makes the selected path assignment structurally invalid. The selected output then feeds the corresponding subcircuit at
the next level of the grid.

At each subcircuit (rules: implication introduction, implication elimination, repetition),
let the activations be $a^{1},\ldots,a^{R}$. Compute
$x=\operatorname{XOR}(a^{1},\ldots,a^{R})$ (true if and only if exactly one rule is active),
mask $b^{i}=x\land a^{i}$, form per-rule outputs $\vec o^{(i)}=b^{i}\land \vec d^{(i)}$,
and output $\vec d^{\mathrm{out}}=\bigvee_{i=1}^{R}\vec o^{(i)}$; thus only the uniquely
active rule (if any) contributes, otherwise $\vec d^{\mathrm{out}}=\vec 0$.

\begin{theorem}[Pointwise evaluator correctness]
\label{thm:section3-correctness}
For every \textbf{DLDS} \(\mathcal D\), goal column \(c\), and path assignment
\(P\in\mathsf{Path}_{N,h}\),
\(\mathrm{acc}_{\mathcal D,c}(P)=1\) iff
\(\mathrm{Invalid}_{\mathcal D}(P)\vee
(\mathrm{WellFormed}_{\mathcal D}(P)\wedge
\mathrm{Discharged}_{\mathcal D,c}(P))\). Consequently,
\(\mathrm{Accept}(\mathcal D,c)\) iff, for all
\(P\in\mathsf{Path}_{N,h}\),
\(\mathrm{Invalid}_{\mathcal D}(P)\vee
(\mathrm{WellFormed}_{\mathcal D}(P)\wedge
\mathrm{Discharged}_{\mathcal D,c}(P))\).
\end{theorem}
The theorem is pointwise in the path assignment \(P\). It characterizes the
operational semantics of the circuit evaluator, not the full Flow-based validity
criterion for compressed \textbf{DLDS}. The relation between this evaluator and
actual \textbf{DLDS} structure is treated in the Lean section: the formalization
proves the evaluator theorem above and also establishes a restricted bridge for
the uncompressed simple-tree fragment.

\section{Lean Formalization}\label{sec:lean}

Lean is an interactive theorem prover based on dependent type theory, designed for machine-checked verification of mathematical proofs and formal systems~\cite{Lean}. It provides a language for defining inductive structures, computable functions, and logical predicates, together with a small trusted kernel that verifies all proofs. We mechanize in Lean the Boolean-circuit evaluator introduced in Section~\ref{sec:Boolean}. The development proceeds from local subcircuit semantics to routing-aware grid evaluation and culminates in the pointwise and global evaluator theorems stated in Theorem~\ref{thm:section3-correctness}. Dependency information is represented throughout by fixed-length Boolean vectors, whose entries indicate whether a given assumption is still active. The formalization also contains \textbf{DLDS}-side structural predicates and a restricted bridge for the uncompressed simple-tree fragment. Thus the Lean development has two layers. The first layer proves correctness of the circuit evaluator for arbitrary path assignments. The second layer starts relating this evaluator back to \textbf{DLDS} structure by proving that, in the simple-tree fragment, a structurally valid \textbf{DLDS} satisfying the route and discharge certificates formalized in Lean yields genuine circuit acceptance of its extracted path. The full compressed case, including the recursive Flow condition for collapsed nodes, ancestor edges, colors, and residual paths, is not mechanized in the present development. 
\paragraph{Local rule semantics.}
The circuit is built from rules grouped into subcircuits, following the structure
described in Section~\ref{sec:Boolean}. In Lean, the local layer is represented
by the objects \texttt{ActivationBits}, \texttt{RuleData}, \texttt{Rule}, and
\texttt{CircuitNode}. An activation bit records whether the premise inputs for a
rule are selected; implication elimination carries two such bits, while
implication introduction and repetition carry one. Each \texttt{Rule} also stores
its rule kind and a \texttt{combine} function computing the output dependency
vector. For implication introduction, this function removes the discharged
assumption; for implication elimination, it takes the bitwise disjunction of the
two premise dependency vectors; and for repetition, it forwards the dependency
vector unchanged. A subcircuit groups alternative rules for a formula. Its Boolean control checks whether exactly one rule is active. This is implemented by the function \texttt{multiple\_xor}, which evaluates to \texttt{true} precisely when a list of Booleans contains a single \texttt{true}. The function \texttt{node\_logic} implements this selection by extracting the activation bits, applying \texttt{multiple\_xor}, masking inactive rules, and
OR-combining the resulting dependency vectors.
The local correctness theorem is \texttt{node\_correct}: if exactly one rule is active, then the node output equals the \texttt{combine} result of that unique active rule applied to the inputs. The proof relies on the equivalence between the Boolean test \texttt{multiple\_xor} and the logical predicate \texttt{exactlyOneActive}. Under this condition, all inactive rules contribute the zero vector, so the final disjunction of outputs reduces to the uniquely active rule output. 
\paragraph{Routing-aware grid evaluation.}
Subcircuits are arranged into layers forming a rectangular grid. Dependency
information is propagated by tokens carrying an origin column, current column,
source column, input label, and dependency vector. A path assignment selects, for
each origin column and level, a target column and input label; target \(0\)
stops the token, while a positive target routes it to the next layer.

At each grid cell, Lean collects the arriving tokens and computes which local
rule is selected by the routing information. The routing-aware evaluator returns
an output dependency vector and an error flag. The lemma
\texttt{node\_logic\_with\_routing\_correct} lifts \texttt{node\_correct} to
this setting: if the routed inputs determine exactly one admissible rule
instance, the evaluator computes that rule and returns no local error; otherwise
the path contributes to \texttt{PathStructurallyInvalid}. Layer errors are
combined by disjunction, and evaluation continues until the goal column is
checked for discharge. Thus the evaluator accepts exactly the two cases used in
Section~\ref{sec:Boolean}: structural invalidity, or no routing error together
with discharge of the goal dependency vector.
\paragraph{Evaluator correctness.}
The function \texttt{evaluateDLDS} constructs the formula-level grid associated
with \(d\), initializes dependency vectors, and runs the circuit evaluator on
the chosen path assignment. The theorem \texttt{dlds\_evaluation\_iff} proves
the operational biconditional of Theorem~\ref{thm:section3-correctness}: for any
\texttt{Graph} \(d\), path assignment, and goal column, evaluation returns
\texttt{true} exactly when the path is structurally invalid, or when it has no
routing error and the goal dependency vector is discharged. The global theorem
\texttt{dlds\_global\_iff} is obtained by universal quantification over path
assignments.
\paragraph{DLDS-side predicates and the simple-tree bridge.} In addition to the evaluator theorems, the Lean development defines structural predicates on \textbf{DLDS} instances. The main predicate is \texttt{ValidDLDS}, which packages the structural and local rule conditions used by the formalized fragment. These conditions include graph hygiene, level and color constraints, simplicity and ancestor-simplicity, the requirement that hypotheses have no incoming deduction edges, local correctness of implication introduction and elimination, root discharge, color acyclicity, ancestor-edge conditions, ancestor-backway information, and non-nested ancestor edges. The predicate \texttt{GenuinelyAccepts} abbreviates the non-vacuous case:
\texttt{PathHasNoRoutingError} together with \texttt{AllAssumptionsDischarged}.
It rules out acceptance caused only by structural invalidity. The bridge currently proved in Lean is restricted to the uncompressed simple-tree fragment. In this fragment, every node has at most one outgoing deduction edge, no node is collapsed, there are no ancestor paths, and formula labels are injective. The theorem states that a valid simple-tree \textbf{DLDS}, together with the executable route and discharge certificates used by the formalization, yields genuine acceptance of the path extracted from the \textbf{DLDS}: \begin{lstlisting}[style=leanstyle] 
theorem tree_bridge_forward 
    (d : Graph) 
    (htree : IsSimpleTreeDLDS d) 
    (hvalid : ValidDLDS d) 
    (hcert : routeCoherentB d = true) 
    (hdis : dischargedB d = true) : GenuinelyAccepts d (pathsFromDLDS d) (goalColumn d) 
\end{lstlisting} 
The variant \texttt{tree\_bridge\_forward\_of\_descent\_coherent} replaces the executable route certificate by a structural layer-coherence assumption,
identifying the remaining obligation needed to remove \texttt{routeCoherentB} in the simple-tree fragment. The discharge certificate remains a separate executable condition in the present development.

The following table summarizes the main Lean objects and the scope of the
corresponding claims:
\begin{center}
\begin{tabular}{p{0.31\linewidth}p{0.34\linewidth}p{0.25\linewidth}}
\toprule
Lean object & Informal meaning & Scope \\
\midrule
\texttt{node\_correct}
  & local node correctness
  & single subcircuit \\
\texttt{circuit\_iff}
  & circuit evaluator equivalence
  & arbitrary grid and path assignment \\
\texttt{dlds\_evaluation\_iff}
  & \textbf{DLDS} evaluator equivalence
  & operational grid semantics \\
\texttt{dlds\_global\_iff}
  & global universal specification
  & all path assignments \\
\texttt{ValidDLDS}
  & structural and local \textbf{DLDS} predicate
  & formalized fragment \\
\texttt{GenuinelyAccepts}
  & no routing error and discharged goal
  & extracted path acceptance \\
\texttt{tree\_bridge\_forward}
  & simple-tree bridge
  & certificate-assisted \\
\texttt{tree\_bridge\_forward\_of\_descent\_coherent}
  & structural routing version of the bridge
  & simple-tree fragment \\
\bottomrule
\end{tabular}
\end{center}

The present Lean development therefore proves the correctness of the circuit
evaluator and a restricted bridge from \textbf{DLDS} structure to circuit
acceptance. It does not mechanize the full recursive Flow condition for
compressed \textbf{DLDS}. Extending the bridge to the full compressed case would
require a separate formal treatment of collapsed nodes, ancestor edges, colors,
dependency bitstrings, and residual paths. We leave this extension for future
work. The complete Lean development, including all definitions, lemmas, and proofs, is
available in the supplementary materials.\footnote{GitHub repository:
\url{https://github.com/lorenzosaraiva/DLDSBooleanCircuit}}
\paragraph{Use of LLM assistance.}
Large language model tools were used during the development of the Lean artifact,
mainly to help draft proof scripts, refactor intermediate lemmas, and identify
missing auxiliary statements. All generated code was reviewed line by line by the
authors, edited where necessary, and checked by Lean.

\section{Related Work} \label{related}

Our mechanized approach for encoding horizontally compressed \textbf{ND} proofs as Boolean circuits benefits from previous works on circuit-based proof verification. We follow the foundational results in~\cite{CookReckhow1979}, which defined propositional proof systems in terms of polynomial-time verifiability by circuits, laying the groundwork for reasoning about proofs as Boolean functions. More recent developments in SAT-based proof checking, particularly in the context of extended Frege systems, as in~\cite{BeamePitassi2001}, also translate logical derivations into circuit representations for efficient validation. 

Additionally, the notion of proof nets in linear logic, as in~\cite{Girard1987}, provides a canonical graphical form of proofs with correctness criteria that are themselves checkable via acyclic switching graphs or circuit-like conditions.
From the perspective of mechanization, several formalizations of proof systems exist in interactive theorem provers. For instance, proof nets and sequent calculi have been mechanized in Coq, notably in the work of Xavier et al~\cite{xavier2018mechanizing}, who formalized focused linear logic in Coq, including cut-elimination and completeness of focusing. \textbf{ND} has been mechanized in Isabelle/HOL, for example through the \emph{NaDeA} assistant~\cite{villadsen2015nadea}, which provides a verified \textbf{ND} system with soundness proved in Isabelle. These efforts focus on formalizing inference systems and meta-theory, whereas our work uses Lean to go one step further: encoding horizontally compressed proofs as Boolean circuits and formally verifying the circuit-based correctness criteria. These works situate our contribution within the broader tradition of circuit-oriented proof theory and mechanized verification. To our knowledge, this is the first Lean mechanization of compressed \textbf{ND} proofs into Boolean circuits, combining compression, modular circuit construction, and machine-checked correctness.

A separate line of work addresses the \emph{size} of dag-like proofs in classical 
and intuitionistic logic. Je\v{r}\'abek~\cite{Jerabek2025} recently showed that 
dag-like natural deduction, in the sense of subformula sharing, does not 
significantly shorten proofs relative to tree-like natural deduction in 
implicational logic. This result concerns a different notion of compression than 
ours: the \textbf{HC} algorithm~\cite{haeusler2025horizontal} collapses nodes at 
the same derivation level and introduces ancestor edges to track provenance, 
producing structures whose size bound follows from the subformula property rather 
than from subformula sharing. Our circuit encoding does not rely on any general 
proof-shortening claim, the $O(N^3)$ circuit size follows from 
$N=|\mathrm{Sub}(\alpha)|$ alone, but the distinction matters for interpreting 
the role of \textbf{DLDS} compression in the broader proof-complexity landscape.

In the classical framework of Cook and Reckhow~\cite{CookReckhow1979},
propositional proof systems are characterized by polynomial-time verifiable
certificates, typically realized by Boolean circuits that check the correctness
of syntactic derivations. Our approach is different: the circuit is not used as
an external checker for a fixed derivation, but as a uniform evaluator for
candidate dependency-propagation paths through the formula-level grid. The
acceptance condition is expressed as a universal property of this evaluator over
path assignments. We emphasize that, in our setting, Boolean values encode
path-selection choices along derivation grids, rather than propositional truth
assignments as in the Cook--Reckhow framework.

\section{Conclusion and Future Work} \label{future}
Our Boolean circuit encoding of compressed \textbf{ND} proofs provides a
foundation for further exploration in both automated and quantum
verification. One promising direction concerns graph non-Hamiltonicity:
for any graph $G$, a formula $\neg \alpha_G^*$ in purely implicational
minimal logic encodes the non-Hamiltonicity of $G$, with a polynomial height-bounded \textbf{ND} proof in the size of $G$, see Corollary 2.5 in~\cite{GH2022}, p.~203. After \textbf{HC}, this yields a \textbf{DLDS} with a polynomial-size kernel\footnote{The polynomial size bound follows from horizontal compression via level-wise merging and subformula-bounded encodings~\cite{haeusler2025horizontal}; Lean formalization at \url{https://github.com/RCMBF/Horizontal-Compression}. This is distinct from standard dag-like proof representations studied by Je\v{r}\'abek~\cite{Jerabek2025}, which concern subformula sharing rather than derivation-level compression.}, which we translate into a Boolean circuit guided by a path array.

The circuit-based formulation of the dependency-propagation evaluator opens a
possible route to quantum verification. Since the pointwise evaluator is a
Boolean circuit and global acceptance is a universal property over path
assignments, future work may study whether the induced circuit behaves as the
constant-$1$ function on the relevant path space. A second direction is extending
the Lean bridge to the full compressed Flow-based \textbf{DLDS} setting.

\bibliographystyle{eptcs}
\bibliography{references}

\end{document}